# Fidelitous Augmentation of Human Accelerometric Data for Deep Learning*

T. K. M. Lee
Singapore Polytechnic, Singapore
School of Information Technology
Monash University Malaysia
tracey_km_lee@ichat.sp.edu.sg

H. W. Chan
Dept of Aerospace Engineering
University of Glasgow
Singapore
2427293C@student.gla.ac.uk

K. H. Leo
Singhealth Office for
Organizational Transformation
Singapore
leo.kee.hao@singhealth.com.sg

E. Chew
Division of Neurology
National University Hospital
Singapore
effie_chew@nuhs.edu.sg

L. Zhao
Division of Neurology
National University Hospital
Singapore
ling_zhao@nuhs.edu.sg

S. Sanei
School of Science & Technology
Nottingham Trent University
United Kingdom
saeid.sanei@ntu.ac.uk

*Abstract*—**Time series (TS) data have consistently been in short supply, yet their demand remains high for training systems in prediction, modeling, classification, and various other applications. Synthesis can serve to expand the sample population, yet it is crucial to maintain the statistical characteristics between the synthesized and the original TS : this ensures consistent sampling of data for both training and testing purposes. However the time domain features of the data may not be maintained. This motivates for our work, the objective which is to preserve the following features in a synthesized TS: its fundamental statistical characteristics and important time domain features like its general shape and prominent transients.**

**In a novel way, we first isolate important TS features into various components using a spectrogram and singular spectrum analysis. The residual signal is then randomized in a way that preserves its statistical properties. These components are then recombined for the synthetic time series. Using accelerometer data in a clinical setting, we use statistical and shape measures to compare our method to others. We show it has higher fidelity to the original signal features, has good diversity and performs better data classification in a deep learning application.**

## I. Introduction

The outstanding achievements in prediction, simulation, and classification have been made possible by deep learning as a component of artificial intelligence. To achieve robustness, enhance generalization, and avoid overfitting of results, deep neural networks (NNs) are trained on large amounts of high dimensional data.

But TS data collection is limited due to various constraints such as privacy, expense, time and physical access. Thus there is a need to augment such data to effectively train machine learning systems. Besides, other uses for synthetic time series are for testing null-hypotheses, validating new methodologies and producing ensembles for event attribution.

Synthesizing a TS requires an understanding of the original data generating process. Thus, for processes in a steady state, it is simple to use the mean and standard deviation of a signal to synthesize the data while adding a small amount of random noise. Together with many other natural phenomena occurring over short time periods these have been successfully modeled by linear stationary stochastic processes. However in TS important transient events occur like mechanical shocks or in exceptional events surrounding process failure.

From these considerations we see that a synthesized TS should exhibit *diversity* — having a useful spread of characteristics so systems can generalize better about the data and have *fidelity* in preserving the i) basic statistical properties of the original as well as ii) important time-based features such as its *shape* and iii) important transients.

In this paper we employ various measures to compare the fidelity of synthesized data with the original. We apply these considerations in a biomedical application to classify human rehabilitative accelerometric data, a common use case [1].

In Section II we describe background material and the motivation for our approach. Section III outlines our physical setup followed by the theory and approach in Section IV. Our experimental results are presented in Section V and we summarize our discussion with conclusions in Section VI. For the rest of the paper we use the terms time series and signal interchangeably and use the notions of fidelity and diversity in an empirical way.

## II. Background Work

In this section we review TS synthesis approaches and ways to preserve their fidelity as well as introduce diversity in their features.

Lately, there has been progress in the generative synthesis of TS using various generative adversarial neural networks (GANs), as seen in [2]. A survey of such syntheses – part of a larger TS augmentation review – in [3] commented that the quality metrics for these type of syntheses are probabilistic in nature and not comparable to traditional augmentation approaches. Also, these methods are resource intensive in terms of training and deployment. In comparison, traditional methods of time series synthesis take up fewer resources and their mechanisms more easily understood. These considerations motivate for our approach. We thus refer to the work by Iwana and Uchida [4] who prioritized some techniques which work well with a variety of classifiers especially those using Convolutional Neural Networks (CNNs). Of these, two widely

Ministry of Education of Singapore grant 2010MOE-IF-005
*An abridged version of this paper appears in the proceedings of IEEE HealthCom 2023
doi.org/10.1109/Healthcom56612.2023.10472398.

used synthesis methods by Le Guennec et al. [5] are window slicing and warping. We note that in these reviews, many time series synthesis methods do not aim to preserve the original features of the data but randomly perturb the time series in empirical ways to generate data.

In earlier works [6][7] we augmented our training data through the use of the surrogate data technique (described in Sec. IV.C) to successfully train a NN-based classifier. But the waveforms of the synthesized time series in many cases were quite unlike the shape of the original signal as seen in Fig 5E. Subsequently we used Singular Spectrum Analysis (SSA) [8] (described in the next subsection) to obtain better results and portions of this work have been used here for continuity.

*Time series feature isolation* To preserve the shape of the original and its important transients we note that SSA non-parametrically extracts trend, cycle and noise (or low level) components of a time series. Thus we advantageously specify the trend, cycle (and seasonal parts) as its *shape*. SSA has been used extensively in biomedical applications [9] and work by Vautard et al. [10] focuses on short, noisy signals which are applicable to our data.

The detection of transients using spectrograms have been well studied [17] and applied.

*Time series feature randomization* Another requirement for the synthesized data is to have the basic statistical features of the original namely: its mean, standard deviation and power spectrum or autocorrelation. These will ensure consistent overall *population* statistics for test and training data. The method of *surrogate data* described by Theiler et al. [11] was designed to preserve these features.

A similar method was used by Kostenko and Vasylyshyn who explored how to improve the effectiveness of spectral signal analysis and in [12] they describe how to generate surrogate data based on the noise component of a signal decomposed by SSA.

*Classification* Using NNs for time series classification have proven very successful, beginning with early approaches such as transforming 1D data to 2D images as the main advances in machine learning involved images. To our knowledge, we were the first to effectively use surrogate data to train systems to classify time series [6]. We used transfer learning from large 2D CNNs with millions of parameters. Subsequently we showed that a 1D CNN and a Long Short-Term Memory (LSTM) NN was able to give very good results with a 50-fold augmentation of data. The 1D CNNs use fewer computing resources and our one hidden layer LSTM network can give better results but its structure is more complex and its training requires more resources. Furthermore, NNs in a many-layered deep configuration have been shown to be more efficient in terms of training and classifier complexity [13].

## III. DATA COLLECTION AND SETUP

Our dataset comprises data from a triaxial accelerometer embedded in a 10 cm cube. The cube is moved by subjects in a manner prescribed by a rehabilitative test. Details of the hardware and test protocol are in [14]. Three channels of data are digitized to 8 bits at a sampling rate of 30 Hz. The movements are visually scored by clinical staff and starts from a score of 3 for completion of the task within 5 seconds with appropriate hand, arm and posture movements. A score of 2 is given when the subject completes the task with great difficulty and/or takes abnormally long time, from 5 to 60 seconds. For a score of 1 which indicates partial completion, the timing would be greater than 60 seconds. Also being able to just grasp, hold and lift the cube would be sufficient to warrant this score. Being unable to do any of these results in a score of 0. In Fig. 1 we see how it is gripped, held vertically and moved.

The data were recorded from 34 patients who have had a history of stroke and undergone rehabilitation. The medical trial was conducted in a hospital over two months. The participants of this study did not give written consent for their data to be shared publicly, so due to the sensitive nature of the research supporting data is not available.

From these subjects, 78 sets of data were recorded. Of these 31 scored at 3, 38 scored 2, 6 scored 1 and 3 scored 0.

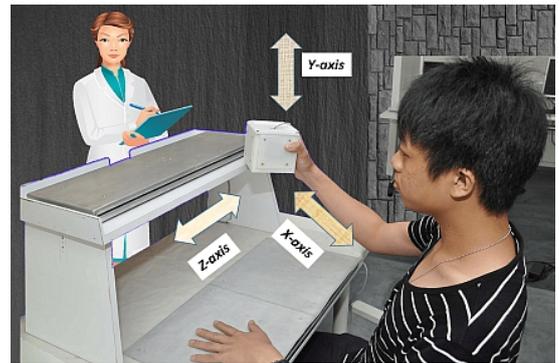

**Fig. 1.** Rehabilitative test with scorer in background.

Thus we consider this a classification exercise where the scores are subjectively given by the scorer. This dataset is challenging due to the lack of visible inter and intra class features as well as visual subjective scoring over time.

*A. Dataset augmentation*

To reduce statistical bias we generate synthetic data so that each class has roughly the same number of time series. We denote various versions of the synthetic dataset by the fold increase of time series being synthesized - so a 10-fold increase would have 10 time series synthesized from an original as shown in Table 1. However data augmentation with scores of 0 and 1 are 12 and 6 times greater respectively, to maintain equitable sample sizes. We retain the fold designator for convenience so the 10-fold for score 3 is actually 60-fold for score 1.

*B. Data preprocessing*

The motion data captured over a session are manually segmented and compensated for noisy readings.

## IV. THEORY AND APPROACH

The theory for the components of our system is covered here concerning transient detection, SSA and surrogate data. In our recent work [15] we showed that our data exhibits linearity. This gives good support for processing the data using SSA and surrogate data which are essentially linear operations.

TABLE I. EXAMPLES OF DATASET AUGMENTATION. QUANTITIES IN THE CENTER CELLS - NUMBER OF TIME SERIES: X-FOLD - FOLD INCREASE. SCORES OF 0 AND 1 ARE AUGMENTED MORE - SEE TEXT.

| Score | Original : # T.S. | "10-fold" # T. S. | "50-fold" # T. S. |
|---|---|---|---|
| 3 | 31 | 310 | 1550 |
| 2 | 38 | 380 | 1900 |
| 1 | 6 | 360 | 1800 |
| 0 | 3 | 360 | 1800 |
| Total | 78 | 1410 | 7550 |

### A. Singular spectrum analysis

SSA is a subspace analysis method originally developed for single channel time series analysis. We describe basic SSA after Vautard et al. [10] where for a time series at time $t$ the data is represented by a vector $\mathbf{x}(t) = \{x(t): t=1... N\}$ with $N$ sequential, equally spaced time intervals. From this a trajectory matrix $\mathbf{Y}$ is formed by sliding a window of length $M < N$ over $x(t)$. The rows of $\mathbf{Y}$ are:

$$\mathbf{x}_1 = [x(1), x(2),..., x(M)]$$
$$\mathbf{x}_2 = [x(2), x(3),..., x_i(M+1)]$$
$$\mathbf{x}_{N-M+1} = [x(N-M+1), x(N-M+2),..., x(N)] \text{ for row } N-M+1$$

and $\quad \mathbf{Y} = [\mathbf{x}_1; \mathbf{x}_2; \cdots ; \mathbf{x}_{N-M+1}]$

where ; denotes vertical concatenation.

The decomposition of $\mathbf{Y}$ into its principal components starts by forming the $M \times M$ covariance matrix

$$\mathbf{C} = \mathbf{Y}^T \mathbf{Y} / N$$

where $^T$ is the transpose operator. Diagonalization of $\mathbf{C}$ produces sorted scalar eigenvalues $\lambda$ and eigenvectors $\mathbf{e}^k$ (length $M$). The singular values of $\mathbf{Y}$ are $\sqrt{\lambda}$, together with $\mathbf{e}^k$ these are used to form principal components (PC), the $k^{th}$ PC is a vector of length $N-M+1$ whose elements are:

$$a_t^k = \sum_{j=1}^{M} x(t+j) e_j^k \quad \text{for } 0 \le t \le N - M \quad (1)$$

In our earlier work [7] we have successfully used $M = 17$. To view the effects of the decomposition, we generate the reconstructed signal component (RC). For a set $K$ of PCs used for reconstruction, the $k^{th}$ RC is a vector of length $N$ given by separate equations catering for the beginning and end conditions of the embedding operation:

$$RC(t) = \frac{1}{M} \sum_{j=1}^{M} \sum_{k \in K} a_{t-j}^k e_j^k \quad \text{for } M \le t \le N - M + 1$$

$$= \frac{1}{t} \sum_{j=1}^{t} \sum_{k \in K} a_{t-j}^k e_j^k \quad \text{for } 1 \le t \le M - 1$$

$$= \frac{1}{N-t+1} \sum_{j=t-N+M}^{M} \sum_{k \in K} a_{t-j}^k e_j^k \quad \text{for } N - M + 2 \le t \le N$$

*Significant eigenvalues* The decomposition returns a set of $M$ eigenvalues which are sorted to show the contribution of an eigenvalue to the variation of the data. We can compute a threshold above which are considered as *significant* eigenvalues Using Relevant Dimension Estimation [16] the corresponding significant *eigenvectors* are used to construct RCs which are considered as trend and seasonal data. The remaining values are then the low level part of the signal.

We have found that SSA does not isolate transient waveforms well, motivating for another method to do so.

### B. Transient detection

For transient detection we use a spectrogram where each pixel in Fig. 2 represents the signal energy present at a given time point and frequency bin. Note this is an example signal from [17] and shows two components: a constant signal *cn* with strong harmonic content and transient signals *tr*. Here *cn* contributes to the horizontal lines or rows showing significant frequency components in bins that last for the signal duration across time. The vertical lines or columns denote significant transients from *tr* in time-bins. In our application, we consider a significant transient as having relatively high energy at short time points *across* the frequency-bins This can be interpreted as the time point (column) having a high mean value and a low standard deviation (*std*). In terms of a spectrogram, we want a uniformly bright narrow strip across its vertical extent.

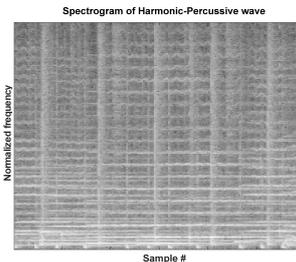
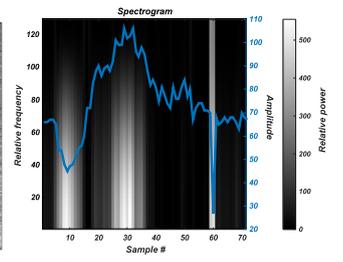

**Fig. 2.** Spectrogram of sample wave with strong harmonics and transients [17].

**Fig. 3.** Spectrogram heatmap of accelerometer. Actual signal superimposed (blue). Strong transient at sample 60.

For example in Fig. 3 we show the time series data of the *y-axis* movement of subject *p18*, trial 3, overlaying the spectrogram. There is a large downward transient at the time point marked by sample #60 depicted by the prominent vertical bar. This is caused by the cube unexpectedly hitting an object and may indicate fatigue at the end of the move. For our signals which are digitized to values of ±128 (Sec. III) we experimentally define a prominent transient as having a mean and standard deviation threshold greater than 50 and 80 respectively in a time point (column). A transient may occupy more than one time point so we need to cluster adjacent time points with similar characteristics. The transients are subtracted from the signal and the intervening data points linearly interpolated, described as follows:

*Transient detection algorithm*
```
1 Generate spectrogram of zero-mean signal
2 For all time points
    compute mean and std
    cluster the means using k-means
3 For all clusters
    remove members that are:
    <  mean threshold   % low energy
    >  std threshold    % energy not evenly spread
    remove clusters with few members
4 For valid clusters  % members adjoining time points
```

```
find start and end values of time points
linearly interpolate start/end times
subtract from actual signal %transient only
store differences % to be restored
```

Finally the *differences* (described in step 4 of algorithm) are added back to the surrogate along with the trend/cycle. We used a window of two time points and 129 normalized frequency points.

### C. Randomizing with surrogate data

In this section we briefly describe the method of surrogate data as well as its output, both which have the same name. We denote the signal by *x* for brevity rather than *x*(*t*). All data sorts mentioned are ascending and bold characters are vectors.

1. Sort *x* to $x_S$ with the indexes of the sorted values $i_x$
   Sort $i_x$ to form the rank index vector $i_{rx}$.
2. Generate a vector of length *N* from a normal distribution.
   Sort this vector to give *rv*.
3. Permute *rv* using $i_{rx}$ as indexes to create a new vector *rrv*.
4. Compute the Fourier transform of *rrv* as *ft*.
   Produce vector *φ* of length *N*/2 of random angles from [0 2π].
   Compute the phase randomized vector $ft_r$:
      For the first half of $ft_r$ multiply this by exp($i\varphi$)
      For the second half of $ft_r$, take the flipped complex conjugate of the first half.
5. Take the inverse Fourier transform of $ft_r$ to form vector *s*.
6. Sort *s* to obtain indexes $i_s$ ; sort $i_s$ to form rank indexes $i_{rs}$
7. Permute $x_s$ using $i_{rs}$ as indexes to get the surrogate vector.

### D. Randomizing with window slicing / warping

Here we briefly describe two window distortion methods. Window slicing discards 10% of data at random starting points of a signal. Window warping expands or contracts a random window of 10% of the data in the signal. In both cases the time series data were interpolated to retain the duration of the signal.

### E. 1D Convolutional neural network

We used a deep 1D CNN as it has shown good results with a relatively simple structure. This had 3 sets of convolutional and pooling layers, then 2 sets of dense and dropout layers, all with ReLU activations. The final dense layer used softmax activation with 3 outputs resulting in a network of 26,800 parameters. We used a mini-batch size of 20 over 20 epochs with the ADAM adaptive weight update rule: this starts from a learning rate of 0.001, decaying at a rate of 1E-6 and uses accuracy as the learning metric. In the augmented dataset, 80% was used for training, 20% for validation and the original 78 time series used for testing.

The input needs to be a fixed length vector which we have set to 91, the median of our variable length data. The time series is zero-padded or truncated to this length before being used to train the 1D CNN.

### F. Methodology

We denote our proposed system as the transient, shape, statistics saving surrogate (TS4) method. The input signal is decomposed into transients, then trends, cycles and low level components by SSA. The transient, trend-cycle and seasonal data or shape of the signal is passed on unchanged and thus exhibits fidelity to the original signal. The low level signal is subjected to surrogate data processing. All of these signal components are recombined to form the synthetic time series as shown in Fig. 4. This process is repeated for the specified fold increase to generate a batch of time series used for training, classification and so on.

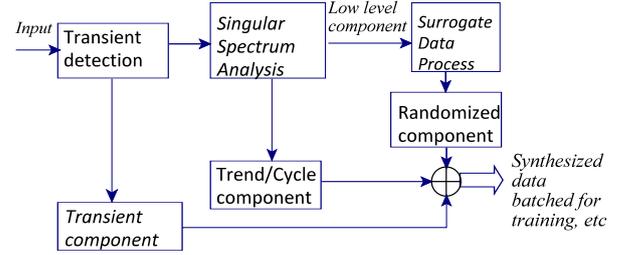

**Fig. 4.** Block diagram of our data augmentation system. Input split by transient detection and SSA into shape(trend/cycle) and low level components. The latter are subject to the surrogate randomizing process. Data is recombined, batched and used to train the classifier (lower right).

For window slicing and warping, the time series data are generated by the software from [4] for the number of fold increases required.

## V. RESULTS

In this section we compare the statistical and shape preserving abilities of our algorithms followed by their effectiveness in classification. Note that the statistical values and signal shapes are only indicative and will differ slightly for each iteration of the generating algorithm.

We use the changes in four numerical measures to compare the preservation of features in the synthesized signals.

i) Signal mean: The difference between the simple average of the original and synthesized signal values as a percent of the average of the original signal.
ii) Signal standard deviation: Same as i) except we compute the standard deviation instead.
iii) Signal autocorrelation function (ACF): the root mean squared difference between the magnitudes of the ACF of the two signals.

The fourth measure gauges the similarity between the signals using:
iv) Dynamic Time Warping [18] (DTW): non-linearly warps two time series to match optimally. This produces a distance *also* called the DTW measuring how closely they match. The signals are normalized to zero mean and unity standard deviation. After computing the DTW this is further normalized by dividing by the length of the signal.

Finally we do a visual comparison in Fig. 5 where the plots of the synthetic time series are overlaid on the original signal. Panel A) shows the original signal with isolated transients, SSA decomposed trend-cycle and low level components. As defined, the *shapes* of the signals are consistent with the gross movement of the cube in Fig. 1. There are some higher frequency components which are a combination of tremors as well as electronic noise [14] which are considered as comprising the low level signal. In Fig. 5B the waveform from our work together with panels C) and D) show a general adherence to the original shape.

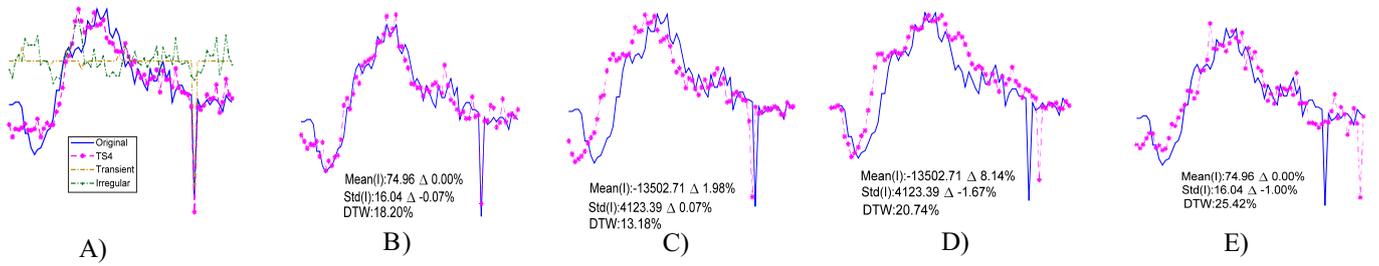

Fig. 5. Examples of waveforms: all solid blue lines are original signal. Overlays are A) starred: SSA derived trend/cycle with transient | dash-dot: low level | dash:transient only B) starred: TS4 C) starred: windows slice D) starred: windows warp E) starred: surrogate only.

In comparison panel E) shows the non-shape preserving surrogate signal from our previous work [7] and the shifted transient. The summary comparison in Table III shows the values of the percent differences in mean, standard deviation, ACF and DTW between the original and synthesized signal.

TABLE III DIFFERENCES OF STATISTICS AND SIMILARITY BETWEEN ORIGINAL AND SYNTHESIZED TIME SERIES.

| Difference | TS4 | Window Slicing | Window Warping | Surrogate only |
|---|---|---|---|---|
| Δ mean % | 0 | 1.98 | -8.14 | 0 |
| Δ std % | -0.07 | 0.07 | -1.67 | -1 |
| Δ autocorr | 0.29 | 0.62 | 0.98 | 0.83 |
| DTW% | 18.2 | 13.18 | 20.74 | 25.42 |

Generally we see that TS4 best preserves the statistical properties as it produces the smallest changes in values. While the DTW (distance) is larger than the window distortion methods, it is an indication that it provides good diversity but still follows the original signal shape in Fig. 5B. In comparison the window distortion methods do not preserve the statistical data as well although they have good similarity measures in the DTW.

We next show the ACF plots of the synthesized signal overlaying the original in Fig. 6. These refer to the summary comparison information in Table III. As can be seen they follow each other quite closely in form with the surrogate methods in panels A) and D) having the best matches. This is in accord with the intent of the surrogate data method.

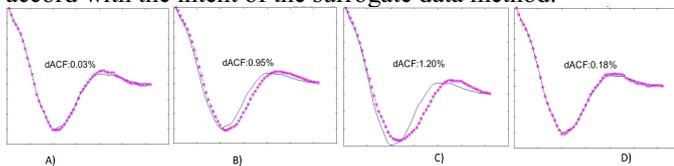

Fig. 6. ACF plots - original signal in solid blue, synthesized in dotted lines. A) TS4 B) Windows Scaling C) Windows Warping D) Surrogate only

Finally we take the synthesized data and perform classification. This was done for each synthesis method and data fold combination, for at least three runs and the median value taken. In Table IV we see our TS4 method gives the best results in being able to predict the condition of a subject with 100% accuracy using just a 25-fold increase of the time series data. The window distortion methods also had good classification together with the surrogate-only method.

## VI. CONCLUSION

In summary we attempted to automatically classify the condition of a subject based on their performance in a rehabilitation test using data from a triaxial accelerometer. The training data are essentially subjective scores awarded by different clinicians over time and not normalized.

By augmenting our data so it has fidelity to the original, we achieve excellent results training a deep 1D CNN to accurately score the movement of subject. Here, much fewer parameters were needed for training, resulting in a smaller network.

Using surrogate data only, our earlier 1D to 2D approach [1] used millions of parameters with 100 fold augmentation to achieve accuracies of around 97%. Another early approach [7] used a 1D CNN with 140,735 parameters and 50 fold augmentation to achieve this.

The key outcome of our work is the generation of synthetic TS that maintains with respect to the original signal its crucial signal characteristics which are: the mean, variance, autocorrelation as well as its shape and transients. These are important in studies of rehabilitative movements involving the picking and placing of objects as well as detecting and preserving important short term events in the move. We show it also performs better than two existing time series augmentation methods.

For greater variety in augmented waveforms the parameters of TS4 can be adjusted. For example, by changing the threshold of what is defined as low level signal in SSA, and the threshold(s) of what is defined as a transient.

TABLE IV ACCURACIES OF FOLD INCREASE WITH FOUR TIME SERIES SYNTHESIS METHODS.

| Synthesis method | TS4 | | | | Window Slicing | | | | Window Warp | | | | Surrogate | | | |
|---|---|---|---|---|---|---|---|---|---|---|---|---|---|---|---|---|
| Fold increase | 50x | 25x | 10x | 5x | 50x | 25x | 10x | 5x | 50x | 25x | 10x | 5x | 50x | 25x | 10x | 5x |
| Predict accuracy | 1 | 1 | 0.933 | 0.892 | 0.987 | 0.987 | 0.933 | 0.933 | 0.96 | 0.853 | 0.933 | 0.973 | 0.893 | 0.822 | 0.776 | 0.781 |
| Validate accuracy | 0.998 | .0.994 | 0.997 | 0.952 | 0.99 | 0.994 | 0.971 | 0.914 | 0.981 | 0.876 | 0.97 | 0.943 | 0.932 | 0.912 | 0.901 | 0.922 |
| Train accuracy | 1 | 1 | 1 | 0.923 | 1 | 0.999 | 1 | 1 | 0.979 | 0.912 | 0.951 | 0.948 | 0.941 | 0.932 | 0.91 | 0.881 |

Also, the transients could be subjected to the window distortion methods without greatly affecting the original statistical properties. Although our method has been used for rehabilitative data only, the results are very encouraging as we have compared it favorably with other widely used time series augmentation methods. Further work would involve other kinds of features to preserve and randomizing schemes to generate more usable synthetic data.

ACKNOWLEDGEMENT

We thank Brian Iwana, Eric Breitenberger, Prof D. Kugumtzis and his team for making their software available. We also thank the anonymous reviewers whose comments have greatly improved the paper.